\documentclass[sigconf,authorversion]{acmart}

\AtBeginDocument{%
  \providecommand\BibTeX{{%
    \normalfont B\kern-0.5em{\scshape i\kern-0.25em b}\kern-0.8em\TeX}}}

\acmConference[]{}{}{}

\usepackage{amsmath}
\usepackage{xspace}
\usepackage{balance}
\usepackage{enumitem}
\usepackage{ifthen}
\usepackage[skip=1pt]{caption}
\usepackage{subcaption}
\usepackage{hyperref}
\usepackage{multirow}
\usepackage{graphicx}
\usepackage{framed}

\setcopyright{none}
\renewcommand\footnotetextcopyrightpermission[1]{} 
{\endMakeFramed}

\newboolean{showcomments}
\setboolean{showcomments}{true}

\ifthenelse{\boolean{showcomments}}
{\newcommand{\nb}[2]{
		\fbox{\bfseries\sffamily\scriptsize#1}
		{\sf\small$\blacktriangleright$\textit{#2}$\blacktriangleleft$}
	}
}
{\newcommand{\nb}[2]{}
}

\newcommand{\ie}{\textit{i.e.,}\xspace}
\newcommand{\eg}{\textit{e.g.,}\xspace}

\newcommand{\etal}{\textit{et al.}\xspace}

\acmSubmissionID{}

\begin{document}

\title{MBSR at Work: Perspectives from an Instructor and Software~Developers}

\author{Simone Romano}
\email{siromano@unisa.it}
\affiliation{
  \institution{University of Salerno}
  \city{Fisciano}
  \state{SA}
  \country{Italy}
}

\author{Alberto Conforti}
\email{alberto.conforti@unito.it}
\affiliation{
  \institution{University of Turin}
  \city{Turin}
  \state{TO}
  \country{Italy}
}

\author{Gloria Guidetti}
\email{gloria.guidetti@unito.it}
\affiliation{
  \institution{University of Turin}
  \city{Turin}
  \state{TO}
  \country{Italy}
}

\author{Sara Viotti}
\email{sara.viotti@unito.it}
\affiliation{
  \institution{University of Turin}
  \city{Turin}
  \state{TO}
  \country{Italy}
}

\author{Rachele Ceschin}
\email{ceschinr@gmail.com}
\affiliation{
  \institution{Nuovo Centro Clinico}
  \city{Turin}
  \state{TO}
  \country{Italy}
}
\author{Giuseppe Scanniello}
\email{gscanniello@unisa.it}
\affiliation{
  \institution{University of Salerno}
  \city{Fisciano}
  \state{SA}
  \country{Italy}
}


\begin{abstract} 
In this paper, we present the preliminary findings from a qualitative study (\ie semi-structured interviews) on how a \textit{Mindfulness-Based Stress Reduction} (\textit{MBSR}) program, carried out in the Software Development (SD) working context, is perceived by the software developers of a multinational company who participated in the MBSR program and by the instructor who led it. MBSR is a deeply personal and experiential practice in helping individuals manage stress, particularly in high-pressure environments such as workplaces, healthcare settings, education, and other demanding professional or personal situations. Although MBSR has been experimented in different working contexts; surprisingly, it has never been studied in the SD working context where there are several stress factors that developers experience (\eg time pressure and uncertainty about the content of a particular task and its outcome). In this respect, qualitative research can generate valuable insights into the application of MBSR in the SD working context that cannot be captured by standardized quantitative measures. Being MBSR instructors and software developers the key stakeholders in delivering an MBSR program in the SD working context, understanding their first-hand experiences can provide a more detailed picture of the investigated phenomenon. The most important takeaway result of our research can be summarized as follows: despite initial skepticism, the developers recognized personal improvements due to the MBSR practice, though the integration of MBSR techniques in the working context remained challenging.

\end{abstract}



\settopmatter{printfolios=true}
\maketitle

\section{Introduction} \label{sec:Introduction}
The software industry is human capital intensive. As such, software companies need to attract and retain talented software developers if they want to maintain a competitive edge in the market and ensure the release of high-quality software products on time. This is why \textit{Big Tech} companies such as \textit{Google} or \textit{Meta} have been investing in \textit{peopleware}~\cite{graziotin2015feelings}, which encompasses anything related to the role of people in Software Development (SD)~\cite{constantine:1995}. 
Peopleware has attracted the attention of the Software Engineering (SE) research community too. This attention is witnessed by the large amount of work on peopleware, which has explored the effects of sleep deprivation and noisy workspaces on developers~\cite{fucci2018need,romano2018effect}, the role of their emotions/moods and personality traits~\cite{graziotin2015feelings,Romano:2023}, and much~more. 

As software companies must continuously release software products of high quality within tight deadlines and budget constraints, developers can be exposed to high levels of stress, burnout, and reduced motivation~\cite{Romano:2024}. This, in turn, can negatively affect job performance, software quality, and developer retention, making the well-being of developers a critical concern for both the software industry and the SE research community. To limit or even prevent these drawbacks, the work on peopleware has shown that it is paramount to act on extrinsic aspects of the job (\eg fun workplaces) but also on intrinsic aspects~\cite{graziotin2015feelings}. Among the intrinsic aspects, \textit{mindfulness} has been emerging as a promising intervention to let developers achieve the best at work. Rooted in Buddhist traditions and spread in the West by Kabat-Zinn~\cite{kabat2003mindfulness}, mindfulness has shown benefits in improving people's well-being~\cite{vonderlin2020mindfulness}. Moreover, it is thought to enhance people's ability to focus on the task at hand and let people be more resilient to extraneous stimuli and avoid mistakes~\cite{glomb2011mindfulness}.
Big tech companies, such as Google, have recognized its potential, implementing mindfulness programs to support their employees~\cite{bernardez2020effects}. However, despite its growing popularity, there is limited empirical evidence on the direct effects of mindfulness on developers' productivity and work-related outcomes~\cite{bernardez2020effects,Bernardez:2023}.

One of the most prominent mindfulness programs is \textit{Mindfulness-Based Stress Reduction} (\textit{MBSR}) ~\cite{kabat2003mindfulness}. It is a manualized and validated group program focused on training the self-regulation of attention and awareness, thereby enhancing voluntary control of mental processes. MBSR consists of eight weekly sessions (2.5 hours each), home assignments, and a full day on retreat in the second half of the program. The sessions include: training and practice in several techniques such as mindfulness meditation, body scanning, and yoga postures; and group discussions and explorations. Empirical evidence has shown the effectiveness of MBSR on employees' well-being in different working contexts~\cite{lomas2018systematic,vonderlin2020mindfulness}, but not in SD. To fill this gap, we started a two-year research project, named \textit{MOOD} (\textit{Mindfulness fOr sOftware Developers})~\cite{MOOD:2024}, funded under the \textit{PRIN} program of the \textit{Italian Ministry for Universities and Research}. The project involves researchers from two fields, SE and \textit{Occupation Health Psychology}, as well as multinational companies operating in the Italian software industry. In this industry paper, we present preliminary findings from a qualitative study exploring how the developers who participated in an MBSR program, as well as the instructor who led it, perceived the experience. The developers involved in the study worked for one of the project's industrial partners. We were interested in both the perspectives of developers and instructors because they can give a more detailed picture of the investigated phenomenon. Moreover, their perspectives can provide valuable insights that cannot be captured by standardized quantitative measures. Finally, the effectiveness of MBSR can depend not only on the content of the program but also on how it is taught by the instructor and integrated into the working context by the developers. A qualitative approach can help uncover contextual influences, difficulties, and strategies to cope with them when delivering an MBSR program in a new working context such as SD.

\section{Related Work} \label{sec:relatedWork}
Empirical evidence has shown the effectiveness of MBSR on employees' well-being in different working contexts~\cite{lomas2018systematic,vonderlin2020mindfulness}, but not in SD. Bernárdez \etal~\cite{bernardez2020effects} conducted a family of experiments with 130 students enrolled in a SE degree course to assess the benefits of practicing mindfulness---no validated program like MBSR was used---in the context of conceptual modeling tasks. The results suggested that, after practicing mindfulness for some weeks, the performance of SE students in conceptual modeling improves. Later, Bernárdez \etal~\cite{Bernardez:2023} conducted an experiment with 51 developers working for a Spanish software company to assess whether practicing mindfulness---again, no validated program like MBSR was used---improves developers' well-being and performance during daily work tasks. The authors observed that practicing mindfulness for several weeks improved developers' well-being while the results on developers' performance were inconclusive. Volobuev \etal~\cite{Volobuev:2021} investigated, through an experiment with 52 developers, whether practicing mindfulness for ten minutes makes a subsequent programming task easier. The authors reported inconclusive results. Penzenstadler \etal~\cite{Penzenstadler:2022} performed two experiments, with 67 participants, to assess the impact of practicing mindfulness on attention awareness, well-being, perceived productivity, and self-efficacy of computer workers (including developers) and students. No validated program like MBSR was used in this research. The authors observed that practicing mindfulness increases attention awareness, well-being, and self-efficacy, while inconclusive results were reported for perceived productivity. Finally, den Heijer \etal~\cite{denHeijer:2017} studied the perceived benefits of a three-minute mindfulness practice in agile stand-up meetings. To do so, the authors conducted an experiment with eight teams from three software companies. 
Significant improvements were reported on some dimensions (\eg meeting effectiveness) for the group performing the practice. 

\textbf{Differences with Past Work.} We used a validated mindfulness program, MBSR, and gathered the lived experience of developers taking part in the program. Moreover, we gathered the perspective of the MBSR instructor. Summing up, our study aims to increase the body of knowledge on mindfulness in the SD context, being complementary to the studies summarized above.

\section{Methodology} \label{sec:studyDesign}
\hspace{\parindent}\ignorespaces
\textbf{Goal.} Our qualitative study seeks to explore the perspectives of MBSR trainees and instructors on the experience of an MBSR program carried out in the SD working context. To that end, we conducted individual interviews with four software developers who participated in an MBSR program and with the instructor who led~it.

\textbf{Context.} The qualitative study presented in this paper is part of a larger investigation, which includes a quantitative study, conducted but not yet published, spanning four periods: \texttt{T0} to \texttt{T3}. At \texttt{T0}, we asked 15 software developers working for one of the project's industrial partners, a multinational company operating in the Italian software industry, to complete a bug-fixing task on a Java program under controlled conditions. Following this, the participants were divided into two groups: the treatment group, which underwent an MBSR program, and the control group, which did not participate in the program. We did not force the developers in any way to take part in the MBSR program. This was to avoid having unmotivated MBSR trainees. Out of 15 participants, six expressed their willingness to take part in the MBSR program. The interviews discussed in this paper were conducted just after the execution of the MBSR program. For the sake of completeness, both the treatment and control groups performed bug-fixing tasks at \texttt{T1} (immediately after the MBSR program), \texttt{T2} (one month after \texttt{T1}), and \texttt{T3} (four months after \texttt{T1}). While \texttt{T1} was needed to assess the immediate benefits of the MBSR intervention, \texttt{T2} and \texttt{T3} focused on evaluating the retention of any effect induced by the MBSR program over time.

\textbf{MBSR.} Before starting the MBSR program, the instructor had a preliminary phone meeting with each of the MBSR trainees (\ie the developers). The (2.5-hour) weekly sessions of the MBSR program were conducted at the offices of the project's industrial partner on Thursdays, while the full-day retreat took place in a meditation room between the sixth and seventh weeks of the program. This change of location allowed the trainees to experience MBSR techniques outside the work environment, fostering a more immersive experience. Throughout all the sessions, the activities strictly followed the original protocol developed by Kabat-Zinn~\cite{kabat2003mindfulness} (see our online appendix~\cite{online:appendix} for details on the content of each section).

\textbf{Interviewees.} Four developers, among those who participated in the MBSR program, were willing to be interviewed. To gather their willingness, the company sent an internal notice describing the purpose of the interview. The developers were not forced to give their willingness in any way; namely, their participation as interviewees was totally voluntary. The developers we interviewed had a seniority of one, three, 14, and 15 years, respectively. Their age ranged from 26 to 38 years. Two interviewees were male, while the other two were female. Regarding their work task, all of them dealt with the development of software systems, as well as the maintenance and evolution thereof. Three of them were also involved in the testing and deployment of software systems. Moreover, two interviewees dealt with the requirement elicitation and analysis, and design of software systems. None of the interviewed developers had practiced mindfulness in the past. Finally, regarding the instructor, she received her certification as an MBSR instructor in 2016. Since then, she has conducted MBSR programs for organizations of both the public and private sectors. Despite nine years of experience as an MBSR instructor, she has never conducted an MBSR program involving a group of software developers.

\textbf{Data Collection and Analysis.} The data collection of our qualitative study is based on the transcripts of five interviews: four with developers who participated in the MBSR program and one with the MBSR instructor who led it. Each interview involved the interviewer and only one interviewee at a time, was conducted in Italian since it was the native language of both the interviewer and interviewees and was audio-recorded. We used semi-structured interviews~\cite{Wohlin:2012}. That is, the questions listed in the interview script were not necessarily asked in order: depending on how the conversation evolved, some questions were asked before others. Semi-structured interviews allow for improvisation and exploration of the investigated phenomenon~\cite{Wohlin:2012}. In other words, in semi-structured interviews, the interview script is a sort of checklist that the interviewer uses to drive the discussion with the interviewee and ensure that all the topics are covered. Our interview script is available in the online appendix~\cite{online:appendix}. The interviews lasted 10 to 20 minutes, with an average of 15 minutes. After transcribing the interviews, we analyzed the transcripts by using thematic analysis, a qualitative analysis method that allows identifying themes within textual data~\cite{Wohlin:2012}. 

\textbf{Limitations.} The interviewees might not have answered truthfully because, for example, they could be afraid of being judged. In this respect, we did our best to make the interviewees feel at ease. The number of interviewees might threaten the validity of the results as well. We cannot exclude that adding further interviewees leads to new themes.  Finally, our findings might not be generalizable to a different population.

\section{Results}\label{sec:results}

\subsection{Developers' Perspective}

\hspace{\parindent}\ignorespaces
\textbf{Prejudices.} The interviewees reported that, before starting the MBSR program, they were biased and skeptical about the effectiveness of the program and thought that it was complex and difficult. 

\textbf{Initial Difficulties.} Some of the interviewees reported that, initially, they experienced the techniques of the MBSR program as odd. Most of the interviewees reported that they initially found the exercises strenuous and difficult. One of them reported that she felt anxious right after the first session because she was not used to focusing on what was going on in her consciousness. To overcome the initial difficulties of getting through the practice or incorporating the techniques into their daily routines, all the interviewees reported that they had found supports that were either personal (\eg not giving up, focusing on the progress, their sense of duty, relying on other participants, or simply enjoying the practices) or contextual (\eg moments of connection with nature, the setting, the fact that the MBSR instructor reassured everyone, the fact that working remotely gave them more freedom to practice at home or the fact that the weekly meeting became a regular occurrence for them during which they could focus only on themselves).

\textbf{Perceived Benefits.} All the interviewees reported that they were benefiting from the MBSR program on a personal level. In particular, they all reported that they noticed an improvement in their way of dealing with situations that they would normally have experienced as difficult or stressful, keeping a clear mind and adopting a different perspective. 
The perceived benefits differed from one another: one of the interviewees reported that he felt a higher level of alertness and ability to focus on things, another reported being more aware of what was going on in his body, while another reported a higher level of discipline. 
All the interviewees said that they were putting at least one of the learned techniques into practice.
Three out of them also reported a practical situation in which they realized that they were behaving differently than they normally would have before practicing MBSR. As for the working context, three interviewees reported that they had significantly more difficulty applying the MBSR techniques in the working context. The reported benefits were generally sparse. Only one interviewee reported that she had managed to apply the MBSR techniques in the working context, which resulted in being better able to cope with her difficulties (\eg the need to deal with numerous requests at once). The interviewee also reported that she was able to feel her physical needs (\ie to pause and take a few steps before getting back to work) and cope better with overwhelming emotions. Also, she observed a higher level of concentration, and the ability to put aside the tendency to multitask and work more efficiently on one thing at a time. All the interviewees reported that they were confident they would improve their ability to apply the MBSR techniques in both the personal and, most importantly, working contexts if they continued practicing them over time.

\textbf{General Difficulties.} When asked what they found difficult about the MBSR program, the interviewees responded differently, emphasizing the individual differences that come into play in such an experience. One of the interviewees reported that he found it difficult to be rigorous and follow the instructor's instructions at home as he reported not being inclined to ``that kind of experience,'' another responded that he found it difficult to follow the program in general, while two interviewees reported that they found only one technique, body scan, not so suitable for them.
Including the practice and the newly learned techniques into their daily routine was what all the participants shared as a difficulty. The lack of time was the most common explanation for that.

\textbf{Behavioral Intention.} All the interviewees were positive about the possibility of repeating this type of experience and would recommend it to a colleague, friend, or family member.

\subsection{Instructor's Perspective.}

\hspace{\parindent}\ignorespaces
\textbf{Group Peculiarities.} When asked about her perception of the implementation of an MBSR program for software developers, the instructor reported that she was unable to identify any significant differences from other groups of people and that the program was overall followed without significant interruption. One particular characteristic identified in the MBSR trainees is that as software developers they are used to working on processes and software. This made it easier for them to explore their mind by viewing it as a processor, as something they could either log in or log out, that they could change and view objectively.

\textbf{Difficulties.} The small size of the group (six developers) diminished the richness of the experience. However, the size of the group allowed them to focus more on their personal experiences and subjective psychological mechanisms.
One of the difficulties reported was that all MBSR trainees had a low level of concentration. This made it necessary to remind them several times during each session to focus on what they were doing. Moreover, the instructor reported the perception that the trainees were not sufficiently informed about what they would be doing during the MBSR program and outside the sessions in their daily lives. Apart from that, all the trainees were engaged and followed her instructions diligently.

\textbf{Perceived Benefits.} The general perception was that all the trainees were curious about the MBSR program and were motivated to invest in it, feeling that it could have a great impact on their daily lives and even on their jobs as software developers. She reported excellent feedback from the trainees with practical examples of perceived benefits in relation to the management of difficult events or relating to other people or their own thought processes. She said that they had the opportunity to experience the effects of mindfulness, but that it is crucial to maintain the practice consistently in the future to maintain and even improve these effects.

\textbf{Suggested Improvements.} 
When asked whether she would have changed anything about this experience, the instructor replied that it would have been helpful to work with an age-diverse group---the ages of the trainees ranged from 28 to 38 years. This could enhance the universality of the experience and make them realize that everyone goes through similar experiences regardless of who they are and what they do. She would also prefer to work with a larger group. Finally, she would take more time for the initial phone meeting and make the conversation a bit~more~structured.

\section{Conclusion} \label{sec:conclusion}
The results of our qualitative study (\ie semi-structured interviews) on MBSR applied in the SD working context suggest that this program has potential benefits in mitigating stress and enhancing focus among software developers. Despite initial skepticism, the developers reported personal improvements due to the MBSR practice, though the integration of MBSR techniques in the working context remained challenging. Based on our findings, software companies could consider the use of mindfulness-based programs like MBSR as part of their employee well-being initiatives, potentially improving job performance and satisfaction, software quality, and developer retention. Evaluating these supposed benefits, from a quantitative point of view, represents a future research direction of our work. Future initiatives could focus on embedding MBSR techniques into specific SD activities, such as code reviews or debugging, and evaluate their impact on several dimensions, such as concentration and decision-making processes.

\section*{Acknowledgements} 
This work has been, in part, funded by the European Union - Next Generation EU, Mission 4, Component 1, under the PRIN program of the Italian Ministry of Universities and Research, project entitled “MOOD–
Mindfulness fOr sOftware Developers” (ID: D53D23008880006).

\balance

\bibliographystyle{ACM-Reference-Format}
\bibliography{bibliography}


\end{document}